\documentclass[sigconf]{acmart}

\usepackage{verbatim}
\usepackage{bm}
\usepackage{multirow}
\usepackage{multicol}
\usepackage{array}
\usepackage{booktabs}
\usepackage{nomencl}
\usepackage{appendix}
\usepackage{amsfonts}
\usepackage{enumitem}

\AtBeginDocument{%
  \providecommand\BibTeX{{%
    \normalfont B\kern-0.5em{\scshape i\kern-0.25em b}\kern-0.8em\TeX}}}

\copyrightyear{2022}
\acmYear{2022}
\setcopyright{acmcopyright}
\acmConference[WWW '22] {Proceedings of the ACM Web Conference 2022}{April 25--29, 2022}{Lyon, France.}
\acmBooktitle{Proceedings of the ACM Web Conference 2022 (WWW '22), April 25--29, 2022, Lyon, France}
\acmPrice{15.00}
\acmISBN{978-1-4503-9096-5/22/04}
\acmDOI{10.1145/XXXXXX.XXXXXX}
\begin{document}

\title{Alleviating Cold-start Problem in CTR Prediction with A Variational Embedding Learning Framework}

\author{Xiaoxiao Xu, Chen Yang, Qian Yu, Zhiwei	Fang, Jiaxing Wang, \\Chaosheng Fan, Yang He, Changping Peng, Zhangang Lin, Jingping Shao}
\affiliation{%
  \institution{Business Growth BU, JD.com}
}
\email{{xuxiaoxiao1,yangchen198,yuqian81,fangzhiwei2,wangjiaxing41}@jd.com}
\email{{fanchaosheng1,landy,pengchangping,linzhangang,shaojingping}@jd.com}

\renewcommand{\shortauthors}{Xiaoxiao Xu, Chen Yang, Qian Yu, Zhiwei Fang, Jiaxing Wang, Chaosheng Fan, et al.}

\begin{abstract}
We propose a general Variational Embedding Learning Framework (VELF) for alleviating the severe cold-start problem in CTR prediction. 
VELF addresses the cold start problem via alleviating over-fits caused by data-sparsity in two ways: learning probabilistic embedding, and incorporating trainable and regularized priors which utilize the rich side information of cold start users and advertisements (Ads). The two techniques are naturally integrated into a variational inference framework, forming an end-to-end training process. 
Abundant empirical tests on benchmark datasets well demonstrate the advantages of our proposed VELF. Besides, extended experiments confirmed that our parameterized and regularized priors provide more generalization capability than traditional fixed priors.
\end{abstract}



\maketitle

\section{Introduction}

% Embedding & MLP
\begin{figure}[t]
\centering
    \begin{minipage}[t]{0.55\linewidth}
      \centering
      \centerline{\includegraphics[width=1\linewidth]{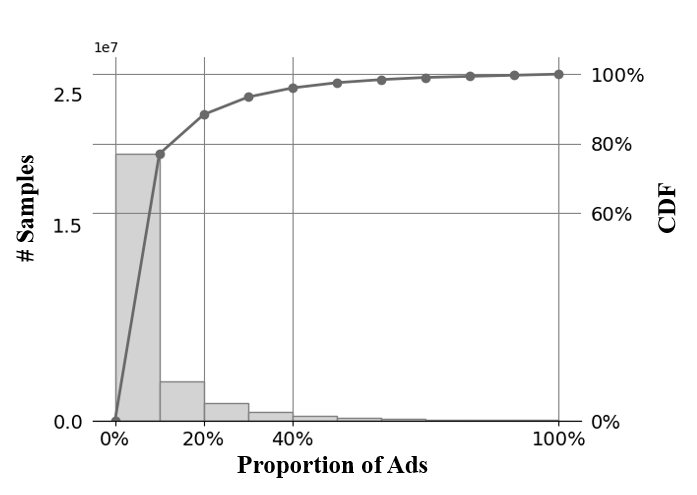}}
      \centerline{\footnotesize{(a)}}
    %   \vspace{-0.5em}
      \centering
    \end{minipage}%
    \hspace{0.2em}
    \begin{minipage}[t]{0.3\linewidth}
      \centering
      \centerline{\includegraphics[width=1\linewidth]{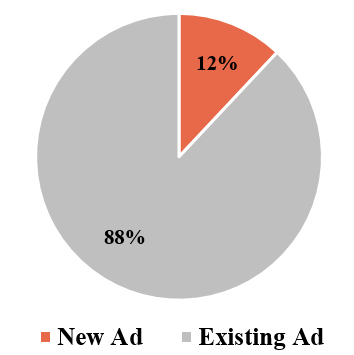}}
      \centerline{\footnotesize{(b)}}
    %   \vspace{-0.5em}
      \centering
    \end{minipage}%
    
    \begin{minipage}[t]{0.55\linewidth}
      \centering
      \centerline{\includegraphics[width=1\linewidth]{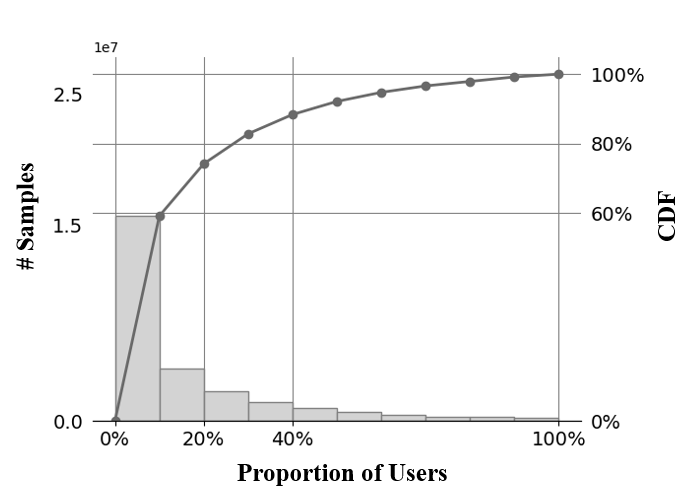}}
      \centerline{\footnotesize{(c)}}
    %   \vspace{-0.5em}
      \centering
    \end{minipage}%
    \hspace{0.2em}
    \begin{minipage}[t]{0.3\linewidth}
      \centering
      \centerline{\includegraphics[width=1\linewidth]{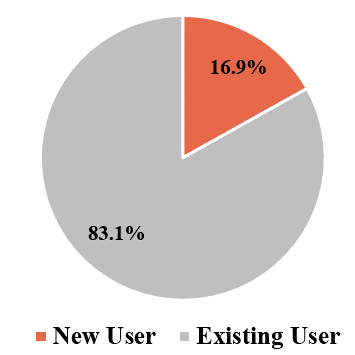}}
      \centerline{\footnotesize{(d)}}
    %   \vspace{-0.5em}
      \centering
    \end{minipage}
    \vspace{-1em}
\caption{Statistics are from Taobao Display Ad Click \protect\footnotemark dataset: The severe long-tail distributions of both (a) Ads and (c) users are shown, and there are above (b) 12\% new Ads and (d) 16.9\% new users in the daily updates.}
\label{new-ID-updating}
\vspace{-3.0em}
\end{figure}

\footnotetext{\url{https://tianchi.aliyun.com/dataset/dataDetail?dataId=56}}

In spite of the impressive development of deep learning in recent decades, cold-start problem keeps becoming a stubborn obstacle in many tasks. Typically, the data scarcity is the major cause of the cold-start problem, and the help by refining model structure is limited. As exemplified by Figure \ref{new-ID-updating}, the severe cold-start problem in online advertising is usually caused by two issues: 1) the drastic long-tail phenomenon which is a widely recognized fact and 2) the remarkable real-time new user and Ad updating.

Being the most successful application of deep neural networks in online advertising, Click-through Rate (CTR) prediction task is also troubled by cold-start problems.

\begin{figure}[t]
\centering
\includegraphics[width=0.8\linewidth]{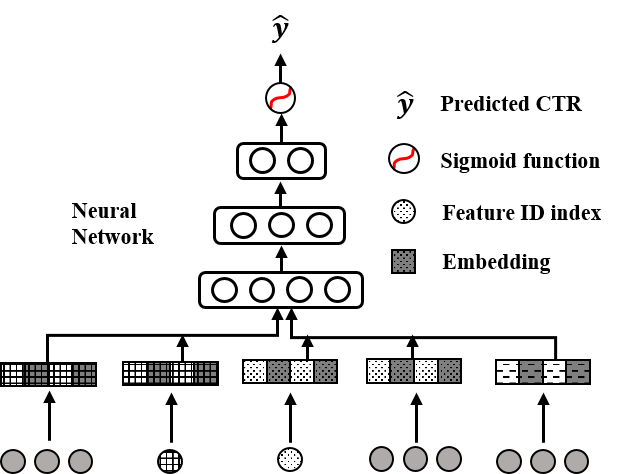}
\caption{An illustration of the traditional Embedding\&Network paradigm structure for CTR prediction.}
\label{CTR-framework}
\vspace{-2.5em}
\end{figure}

The state-of-the-art deep CTR models mostly adopt an Embedding$\&$Network paradigm as illustrated in Figure \ref{CTR-framework}, in which Embedding module works as a representative mapper \cite{liu2018field,guo2021embedding,guo2021dual}. 
Embedding module maps each distinct feature value to a low-dimensional dense embedding vector, where discrete features cover all the categorical features and discretized numerical features. 
The number of trainable parameters in deep CTR models is heavily concentrated in Embedding module, and the Embedding module determines the input distribution of the subsequent feature interaction module and MLP module.
It is considerably data demanding to train a good Embedding module, and this makes it a challenging task in recommendation systems to provide reasonable embeddings for users and Ads with few or no support samples \cite{meta-2}. 
Therefore, enhancing the generalization ability and robustness of the Embedding Module is critical for alleviating the severe cold-start problem in CTR prediction.

Obtaining reasonable embedding for cold-start user and Ad is a challenging task and has been an active research area \cite{volkovs2017dropoutnet,roy2016latent,seroussi2011personalised,content-based-1,content-based-2,mo2015image,meta-1,meta-2,meta-3,meta-4,meta-5}. The existing work mostly concentrates on two kinds of methods, i.e., content-based methods \cite{volkovs2017dropoutnet,roy2016latent,seroussi2011personalised,content-based-1,content-based-2,mo2015image} and meta-learning involved methods \cite{meta-1,meta-2,meta-3,meta-4,meta-5}. Content-based methods introduce richer attributes of user or Ad to obtain a more robust embedding for user or Ad. Meta-learning is involved to transfer knowledge from other users or Ads to the cold-start ones by carefully designed training procedure and sample partition. Both content-based and meta-learning involved methods have been confirmed to be effective. However, these methods are all based on point estimate, i.e., trying to locate a reliable single point in the embedding space for each user and Ad. Previous research has shown that point estimate has a huge risk to result in isolated and unreliable embedding for cold-start user and Ad because of the scarce training samples \cite{zhang2019variational}. In addition, the model for embedding point estimation is prone to overfit unless carefully designed regularization is equipped for parameter tuning \cite{salakhutdinov2008bayesian}.

To make better use of the limited data to obtain more reliable embedding for cold-start users and Ads and avoid overfitting, we propose a general Variational Embedding Learning Framework (VELF). VELF regards the embedding learning as distribution estimate instead of point estimate. 
Through building a probabilistic embedding framework based on Bayesian inference, the statistical strength among users and Ads can be shared, especially by the cold-start ones. 
Bayesian approach has been approved to be more robust regarding to data scarcity \cite{liang2018variational} and more interpretable. 
The distributions are estimated via variational inference (VI) which can avoid the computational intractability. 
To better share the global and statistical strength among users and Ads, we propose to parameterize priors with neural network and the attributes of users and Ads as inputs. 
A regularized and parameterized prior framework combining the parameterized prior with the fixed standard normal prior is proposed to further avoid overfitting.
VI solves distribution estimate as an optimization problem, thus the parameters of the probabilistic embedding framework and the following discriminative CTR prediction network are jointly learned enjoying an end-to-end manner.

In this paper we focus on CTR prediction in the scenario of display advertising. Methods discussed here can be applied in similar scenarios suffering cold-start problems, such as personalized recommendation, sponsored search, etc.
The major contributions in this paper are:
\begin{itemize}[leftmargin=10pt]
\setlength{\itemsep}{-1pt}
    \item We propose a general Variational Embedding Learning Framework (VELF) with an interpretable probabilistic embedding generation process to alleviate the cold-start problem in CTR prediciton. The embedding distributions and the discriminative CTR predicition network parameters are learned end-to-end.
    \item Novel parameterized and regularized priors naturally utilizing the rich side information are designed to further improve the generalization ability of our model.
    \item Extensive experiments are conducted on three benchmark datasets. Results verify the effectiveness of our proposed VELF and the superiorities of proposed parameterized and regularized priors.
\end{itemize}

\section{Related Work}
In this section, we will introduce the related work from two aspects: Cold-start Recommendation and Variational Inference in Recommendation.

\textbf{Cold-start Recommendation: }Cold-start problem is usually caused by two issues: the prevailing long-tail phenomenon and the continuous update of new users and Ads. It is challenging to make recommendations for cold-start users and Ads because of the data limitation. 
In this paper, we focus on the cold-start problem alleviation in CTR prediction task by embedding optimization.

To alleviate cold-start problem in CTR prediction by improving embedding generalizaiton ability, Content-based methods and meta-learning involved methods fall into the same domain with our work. Content-based methods introduce side information for cold-start IDs, i.e., using user and item attributes\cite{volkovs2017dropoutnet,roy2016latent,seroussi2011personalised,content-based-1,content-based-2,mo2015image}, and relational data\cite{lin2013addressing,zhao2015connecting}. DropoutNet\cite{volkovs2017dropoutnet} is the representative of the content-based methods. They can improve the cold-start performance comparing to classic methods that do not use those rich features. However, they do not improve user ID and Ad ID directly. 
Meta-learning is intent to learn the general knowledge across similar learning tasks, so as to rapidly adapt to new tasks based on a few examples. Meta-learning methods have been widely proposed for cold-start recommendation, e.g., learning a meta-learner to better initialize CTR prediciton models \cite{meta-1,meta-2}, utilizing Meta-Embedding \cite{meta-3,meta-4}, in which MWUF \cite{meta-4} is the State-of-the-art. These methods have been validated to be effective, but they need carefully tuning with carefully designed training procedures. 

Above all, content-based methods and meta-learning involved methods are all based on point estimate which still has a huge risk to result in isolated and unreliable embedding for cold-start user and Ad \cite{zhang2019variational}. In addition, the model for embedding point estimate is prone to overfit \cite{salakhutdinov2008bayesian}.

\textbf{Variational Inference in Recommendation: }
Variational inference (VI) has been applied in recommendation but coupling with AutoEncoders, i.e., Variaitonal AutoEncoders (VAEs) \cite{kingma2013auto}. In recommendation, VAEs concentrate on Collaborative filtering and are trying to model the uncertainty of users and items representations, and then collectively reconstructing and predicting user preferences \cite{liang2018variational,askari2020joint,gupta2018hybrid,shenbin2020recvae}. Different from these methods, we apply VI to discriminative models, i.e., CTR prediction task, to alleviate cold-start problem. In our work, VI is the technique we choose to avoid the computationally intractable problem in distribution estimate for users and items embedding by Bayesian inference.

\section{Method}
In Section 3.1, we first review the background of CTR prediction, the fundamentals of Variational Inference and the cold-start issues of Point Estimate. Then we describe Distribution Estimate in our proposed variational embedding learning framework in Section 3.2. We further dig into the details of the implementations during training in Section 3.3 and predicting in Section 3.4. The notations are summarized in Table \ref{tab:notation}.

\subsection{Preliminaries}
\subsubsection{CTR Prediction Problem Formulation}
Given an user, a candidate Ad and the contexts in an impression scenario, CTR prediction, is to infer the probability of a click event. The CTR prediction model is mostly formulated as a supervised logistic regression task and trained with an i.i.d. dataset $\mathcal{D}$ collected from historic impressions. Each instance $(\bm{x}, y) \in \mathcal{D}$ contains the features $\bm{x}$ implying the information of ${\{user, Ad, contexts\}}$, and the label ${y \in \{0, 1\}}$ observed from user implicit feedback. Let $u$ denotes the user ID index, $i$ denotes the Ad ID index, $c(u)$ and $c(i)$ represents the course features, i.e., attributes of $u$ and $i$, the instance features $\bm{x}$ can be expressed as:
\begin{equation}
\bm{x}=[u, c(u), i, c(i), contexts]
\end{equation}
where contexts contain the scene information such as the positions, time, etc.
\begin{table}[]
    \centering
    \begin{tabular}{l|l}
    \toprule
    \specialrule{0em}{1pt}{1pt}
    %\specialrule{0em}{1pt}{1pt}
        %$\bm{x}$ & feature vector of input instance\\
        %$y$ & observed label for input instance\\
        %$\hat{y}$ & estimated CTR for input instance\\
        %$\bm{z}$ & the input embedding of MLP module\\
        $u$, $z^u$ & user id and its embedding \\
        $i$, $z^i$ & Ad id and its embedding\\
        $c(\cdot)$ & the number of attributes of ID\\
        $g_{\bm{\phi}}$ & parameterized function of embedding module\\
        $f_{\bm{\theta}}$ & parameterized function of MLP module\\
        $q_{\bm{\phi}_q}(\bm{z}|\bm{x})$ & approximated posterior distributation\\
        $p_{\bm{\phi}_p}$ & prior distributions of embedding distribution\\
    \bottomrule
    \end{tabular}
    \caption{Important notations.}
    \label{tab:notation}
    \vspace{-2.5em}
\end{table}

With the measurable progress of the research and application of neural network, most recent CTR prediction models share an Embedding and Multi-Layer Perceptron (MLP) paradigm. Specifically in each instance, $\bm{x} \in \mathbb{N}^{m}$ is the vector of feature ID indexes, and $m$ means the total number of selected features. Because the selected features are ID indexes, they have to be encoded into real-value to apply optimization methods. Embedding Module solves this problem by mapping these ID indexes into low-dimensional representations and concatenating them to form the input of MLP Module afterwards, i.e.,
\begin{equation}
\bm{z}=g_{\bm{\phi}}(\bm{x})
\end{equation}
where $g_{\bm{\phi}}(\cdot)$ refers to the function of the Embedding Module, and let $\bm{\phi}$ denotes its parameters. Subsequently, the estimated CTR $\hat{y}$ can be obtained by the following discriminative model,
\begin{equation}
\hat{y}=\sigma(f_{\bm{\theta}}(\bm{z}))
\label{predict}
\end{equation}
where $f_{\bm{\theta}}(\cdot)$ refers to the function of the MLP Module which is parameterized by $\bm{\theta}$, and $\sigma(\cdot)$ is the sigmoid activation function. 
The model parameters $\bm{\phi}$ and $\bm{\theta}$ are learned by maximizing the objective function $\mathcal{L}(\bm{\phi},\bm{\theta})$ with gradient-based optimization methods.
In the traditional point estimate CTR prediction model, the objective function is equal to the negative log-likelihood $l(\bm{\phi},\bm{\theta})$:
\begin{equation}
\begin{aligned}
\mathcal{L}(\bm{\phi},\bm{\theta}) &= l(\bm{\phi},\bm{\theta}) \\
&\equiv -y {\rm log}\hat{y}-(1-y) {\rm log}(1-\hat{y})
\end{aligned}
\end{equation}
% Describe the traditional format here.

\subsubsection{Variational Inference}
Variational inference is an analytical approximation technique to learn posterior distribution $p(\bm{z}|\bm{x})$ of latent variable $\bm{z}$ conditional on the observed variable $\bm{x}$.
Variational inference can work well with deep learning by formulating the Bayesian inference problem in deep learning as an optimization-based approach. Thus, the stochastic gradient descent optimization methods can be adopted.

Now we summarize the fundamentals of variational inference. It is clear the posterior can be formulated as $p(\bm{z}|\bm{x})=p(\bm{x},\bm{z})/p(\bm{x})$ based on Bayes rule. 
However, the marginal likelihood $p(\bm{x})=\int{p(\bm{x},\bm{z})d\bm{z}}$ has no analytic solution or efficient estimator. 
To avoid the computational intractability, variational inference obtain the best approximate posterior distributation $q_{\bm{\phi}_q}(\bm{z}|\bm{x}) \approx p(\bm{z}|\bm{x})$ 
by maximizing the Evidence Lower Bound(ELBO) with respect to the variational parameters $\bm{\phi}_q$:
\begin{equation}
\begin{aligned}
ELBO(\bm{\phi}_q)
&=\mathbb{E}({\rm log}p(\bm{x}|\bm{z}))-D_{KL}(q_{\bm{\phi}_q}(\bm{z}|\bm{x})||p(\bm{z}))
\end{aligned}
\label{Eq7}
\end{equation}
%With Gaussian distribution assumption, the $ELBO$ can be evaluated in closed form.

\subsubsection{Cold-start Issues of Point Estimate}
In prior point estimate methods, Embedding module explicitly maps user ID and Ad ID into a low-dimension embedding space where the similar IDs are expected to be close. However, the embedding points for cold-start users and Ads tend to be isolated because of the ubiquitous data sparsity problem.
To alleviate the probelm, in addition to the only similarity objective supervision introduced by collaborative filtering mechanism with interactions, the existing achievements are enlighten to use content-based and meta-learning based methods with IDs' attributes.

However, there are two important issues: 1) The attributes of users and Ads are only exploited to be as the ID point initialization before training or the fixed final representation of the ID point for inference. Thus the embedding of cold-start user or Ad is still risky to be isolated during training. 2) The point estimate suffers the overfitting problem.
To further alleviate those problems, we are motivated to concentrate on estimating distributions for each user ID $u$ and Ad ID $i$, which exploits the global knowledge during end-to-end training and is much more interpretable. Also, 
it is confirmed in our work that our proposed distribution estimate method is more effective and robust than point estimate methods when data is limited.

\begin{figure*}[t]
\centering
\includegraphics[width=0.9\linewidth]{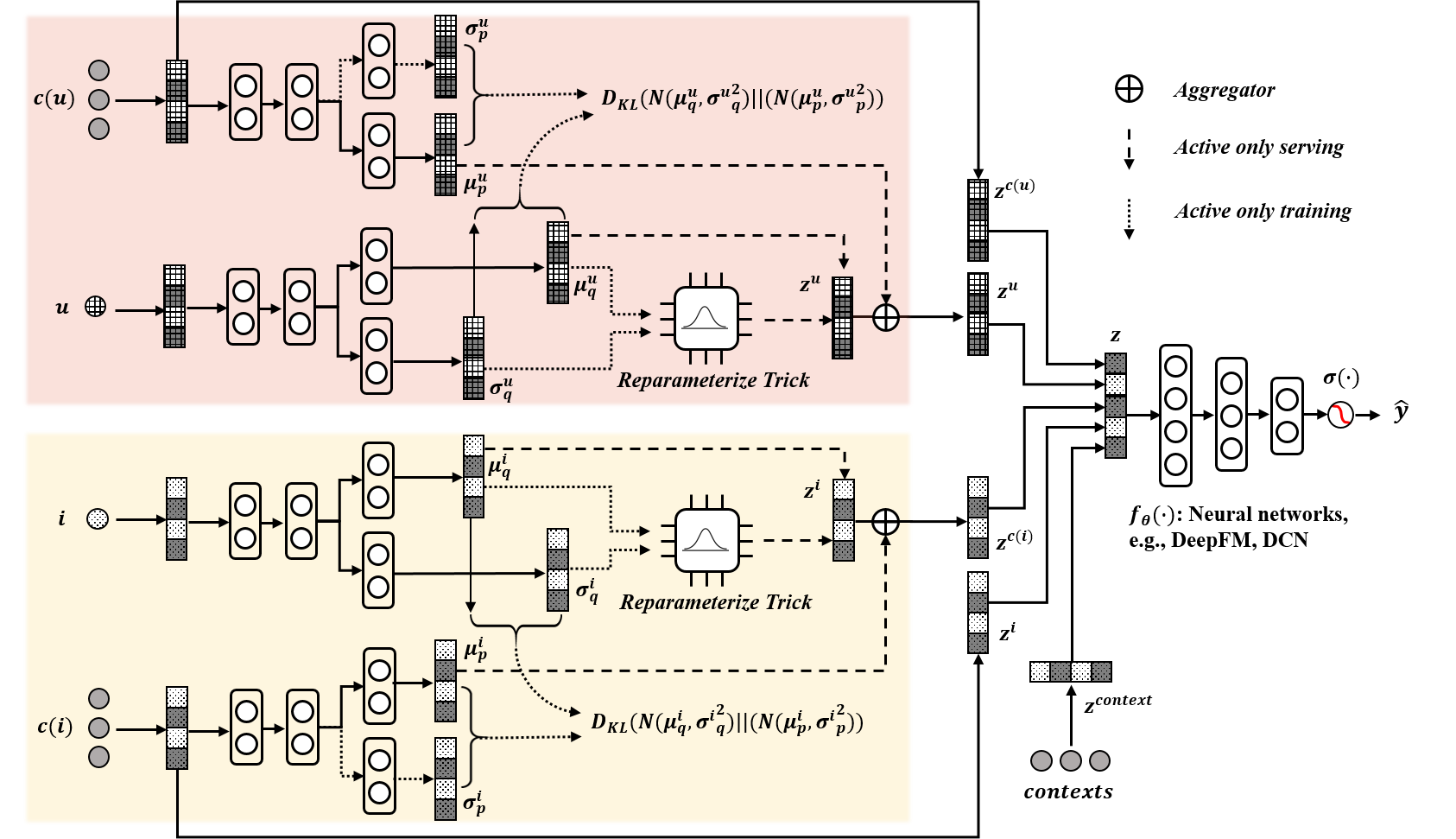}
\caption{An illustration of the variational embedding learning framework for CTR prediction.}
\label{variational-framework}
\vspace{-1.5em}
\end{figure*}

\subsection{Distribution Estimate}
In this section, we focus on addressing the theoretical essentials of our proposed methods, the implementation details will be covered in Section 3.3 and 3.4.
\subsubsection{Variational Embedding Framework in CTR Prediction}
We propose Variaitonal Embedding Learning Framework (VELF) aiming to predict the distribution of each user and Ad embedding. In this way, the model to be learned is regarded as $p_{\bm{\phi},\bm{\theta}}(y|\bm{x},\bm{z})$, and $\bm{z}$ denotes the unobserved latent variables, i.e., embedding space, and the distribution that has to be estimated is the posterior of $\bm{z}$ which is termed as $p(\bm{z}|\bm{x})$. We adopt Gaussian assumptions for all the distributions in our VELF.

Variational inference (VI) is chosen to obtain the approximate posterior distributation $q_{\bm{\phi}_q}(\bm{z}|\bm{x}) \approx p(\bm{z}|\bm{x})$, because it is efficient to be parameterized and computed by neural network. 
According to Section 3.1.2, it is clear that VI casts distribution estimate for latent variables $\bm{z}$ as an optimization problem. Parameterizing the probabilistic models in VI by neural networks, the computationally scalable stochastic gradient-based optimizing methods can be applied \cite{kingma2019introduction}.

In VELF, our objective function is naturally equal to the Evidence Lower Bound (ELBO):
\begin{equation}
\setlength{\abovedisplayskip}{1pt}
\begin{aligned}
\mathcal{L}(\bm{\phi}_q,\bm{\theta}) &= ELBO(\bm{\phi}_q,\bm{\theta}) \\
&\equiv \mathbb{E}({\rm log}p(\bm{x}|\bm{z}))-D_{KL}(q_{\bm{\phi}_q}(\bm{z}|\bm{x})||p(\bm{z}))
\label{Eq7-2}
\end{aligned}
\setlength{\belowdisplayskip}{1pt}
\end{equation}
According to Equation \ref{Eq7-2}, our optimization goal includes two terms. The first term tries to maximize the likehood to improve the confidence of prediction, and the second term tries to find the approximate posterior distribution by minimizing the KL divergence. For binary dataset $\mathcal{D}$ in our CTR prediction scenario, the confidence of prediction $\mathbb{E}({\rm log}p(\bm{x}|\bm{z}))$ is calculated as Log-loss $l(\bm{\phi},\bm{\theta})$ \cite{kingma2013auto}. Adopting the same perspective as \cite{liang2018variational} that the KL divergence term can be viewed as regularization, we introduce a parameter $\alpha$ to control the trade-off between how well the model can fit the data and how close the approximate posterior is to the prior $p(\bm{z})$ during training. To reduce the time-consuming for selecting $\alpha$, we also adopt the annealing method similar to \cite{liang2018variational}: we start with $\alpha$=0, and gradually increase $\alpha$ to 1.
\begin{comment}
\begin{equation}
\begin{aligned}
ELBO(\bm{\phi},\bm{\theta})
&=l(\bm{\phi},\bm{\theta})-\alpha \cdot  D_{KL}(q_{\bm{\phi}_q}(\bm{z}|\bm{x})||p(\bm{z}))
\end{aligned}
\label{Eq8}
\end{equation}
where $\bm{\phi}$ is the whole parameters in Embedding module with $\bm{\phi}_q$ as a part.  
\end{comment}

Now, let us cover the details of $p(\bm{z})$ which is another crux of our method.
$p(\bm{z})$ represents the prior distribution of the latent embedding $\bm{z}$ in Bayes Learning, which is mostly assigned to a certain normal Gaussian distribution. 
We argue that fixed prior limits the generalization capability of our method due to the huge discrepancy among dissimilar users and Ads. 
In our proposed method, we parameterize $p(\bm{z})$ as $p_{\bm{\phi}_p}(\bm{z}|\bm{c})$ by neural networks with the coarse features $\bm{c}$ of user ID and Ad ID as inputs,
where $\bm{\phi}_p$ denotes the specified neural networks parameters.

In this way, we can make full use of the information in dataset to obtain reasonable priors. IDs with the similar attributes can naturally cluster together in the latent embedding space, because they are sampled from similar distributions which are restricted to stay close to the similar prior distributions by the KL divergence regularization. 
Thus, the global knowledge in each cluster can be shared by cold-start IDs with few samples that it contains. With the global knowledge, even the cold-start IDs can obtain reasonable embeddings.

Finally, our maximization goal can be rewritten as:
\begin{equation}
\setlength{\abovedisplayskip}{1pt}
\begin{aligned}
\mathcal{L}(\bm{\phi},\bm{\theta})
&=l(\bm{\phi},\bm{\theta})-\alpha \cdot  D_{KL}(q_{\bm{\phi}_q}(\bm{z}|\bm{x})||p_{\bm{\phi}_p}(\bm{z}))
\end{aligned}
\label{Eq9}
\setlength{\belowdisplayskip}{1pt}
\end{equation}
where $\bm{\phi}=[\bm{\phi}_q, \bm{\phi}_p]$.

\subsubsection{Mean-field Variational Embedding Framework}
In this paper, we aim to alleviate the cold-start problem both for users and Ads . There are two different latent variables, i.e., user latent embedding $\bm{z}^u$ and Ad latent embedding $\bm{z}^i$. On the strength of Mean-field Theory \cite{blei2017variational}, we suppose $\bm{z}^u$ and $\bm{z}^i$ are mutually independent and each governed by distinct factors in the variational density. Then our maximization goal turns into:
\begin{equation}
\begin{aligned}
\mathcal{L}(\bm{\phi},\bm{\theta})
&=l(\bm{\phi},\bm{\theta})\\
-\alpha \cdot (&D_{KL}(q_{\bm{\phi}_q^u}(\bm{z}^u|u)||p_{\bm{\phi}_p^u}(\bm{z}^u))\\
&+ D_{KL}(q_{\bm{\phi}_q^i}(\bm{z}^i|i)||p_{\bm{\phi}_p^i}(\bm{z}^i)))
\end{aligned}
\label{Eq10}
\end{equation}
where $\bm{\phi}=[\bm{\phi}_q^u, \bm{\phi}_q^i, \bm{\phi}_p^u, \bm{\phi}_p^i]$.

\subsubsection{Regularized Priors}
As mentioned before, we introduce unfixed parameterized priors for ID with ID's features as inputs. 
Thus we can make full use of the information in dataset to obtain reasonable priors and facilitate the knowledge sharing among IDs with similar attributes. 
However, the parameterized prior technique still has a risk of overfitting by introducing additional parameters of distributions. To mitigate the overfitting risk, we propose to regularize priors by forcing the parameterized priors to be close to a standard normal hyper-prior: 
\begin{equation}
\begin{aligned}
&p(\bm{z}^u)=\mathcal{N}(0,\bm{I}^u)\\
&p(\bm{z}^i)=\mathcal{N}(0,\bm{I}^i)
\end{aligned}
\label{Eq9}
\end{equation}
With an additional component to the KL-divergence term, We can rewrite our objective function as:
\begin{equation}
\begin{aligned}
\mathcal{L}(\bm{\phi}^u,\bm{\phi}^i,\bm{\theta})
&=l(\bm{\phi}^u,\bm{\phi}^i,\bm{\theta})\\
-\alpha \cdot (&D_{KL}(q_{\bm{\phi}_q^u}(\bm{z}^u|\bm{x}^u)||p_{\bm{\phi}_p^u}(\bm{z}^u)) + \\ &D_{KL}(q_{\bm{\phi}_q^i}(\bm{z}^i|\bm{x}^i)||p_{\bm{\phi}_p^i}(\bm{z}^i)))\\
- \alpha \cdot (&D_{KL}(p_{\bm{\phi}_p^u}(\bm{z}^u)||p(\bm{z}^u)) + \\ &D_{KL}(p_{\bm{\phi}_p^i}(\bm{z}^i)||p(\bm{z}^i))) 
\end{aligned}
\label{Eq13}
\end{equation}

\subsection{Training with Distribution}
In this chapter, we describe the training implementation details of VELF, which is illustrated in Figure \ref{variational-framework}.

%\subsubsection{Probabilistic Encoder}
Here, we take the obtaining of user embedding $\bm{z}^u$ as an illustration, and Ad embedding $\bm{z}^i$ can be obtained in the same way.
With parameterization by neural network, the posterior distributions and prior distributions can be obtained by data-dependent functions:
\begin{equation}
 q_{\bm{\phi}_q^u}(\bm{z}^u|u)=\mathcal{N}(\mu_q^u(u),{\sigma_q^u}^2(u)) 
\end{equation}
\begin{equation}
 p_{\bm{\phi}_p^u}(\bm{z}^u) = p_{\bm{\phi}_p^u}(\bm{z}^u|c(u))=\mathcal{N}(\mu_p^u(c(u)),{\sigma_p^u}^2(c(u)))
\end{equation}
As shown in Figure \ref{variational-framework}, the posterior distribution parameters $\mu_q^u$ and $\sigma_q^u$ are computed from feature ID $u$ with DNNs, and prior distribution parameters $\mu_p^u$ and $\sigma_p^u$ are computed from the attributes of $u$ with DNNs.

%\subsubsection{The Reparameterization Trick}
In VELF, the latent embedding of user ID $\bm{z}^u$ are sampled from the estimated posteriors via reparameterization trick. Given an instance $(\bm{x}, y)$, the resulting user embedding for each sampling can be calculated as:
\begin{equation}
\begin{aligned}
&\bm{z}^u=\mu_q(u)+\sigma_q(u)\odot\bm{\epsilon}^u\\
&\bm{\epsilon}^u \sim \mathcal{N}(0,\bm{I})
\end{aligned}
\end{equation}

As illustrated in Figure \ref{variational-framework}, the embedding of user, Ad, context, and attributes of user and Ad, are concatenated to obtain the input embedding $z$ for the discriminative model:
\begin{equation}
\hat{y}=\sigma(f_{\bm{\theta}}({\rm concat}(\bm{z}^u,\bm{z}^i,\bm{z}^{c(u)},\bm{z}^{c(i)},\bm{z}^{context})))
\label{predict}
\end{equation}
The resulting Log-loss for sample $(\bm{x}, y)$ can be obtained by:
\begin{equation}
l(\bm{\phi},\bm{\theta})=\frac{1}{L}\sum_{k=1}^L (-y {\rm log}\hat{y}_{(k)}-(1-y) {\rm log}(1-\hat{y}_{(k)}))
\label{Eq16}
\end{equation}
where each $\hat{y}_{(k)}$ is calculated by Equation \ref{predict} with randomly sampled $\bm{\epsilon}^u$ and $\bm{\epsilon}^i$, and $L$ is the total Monte Carlo sampling number for each instance and is fixed to be 1 in this paper. 
The KL-divergence terms in $ELBO$ can be computed and differentiated without estimation following the given definition for Gaussian distributions:
\begin{equation}
D_{KL}(q||p)={\rm log}\frac{\sigma_p}{\sigma_q}+\frac{\sigma_q^2+(\mu_q-\mu_p)^2}{2\sigma_p^2}
\label{Eq17}
\end{equation}
We apply Equation \ref{Eq16} and Equation \ref{Eq17} to Equation \ref{Eq13}, yielding the differentiable $ELBO$, i.e., our objective function. By maximizing the differentiable $ELBO$, the variational parameters $\bm{\phi}$ and the discriminative model parameters $\bm{\theta}$ are jointly learned in an end-to-end manner. 
The computationally scalable mini-batch stochastic gradient descent methods are adopted during the training procedure.

\subsection{Predicting with Distribution}
As shown in Figure \ref{variational-framework}, given a VELF model trained following Section 3.3, we make predictions using the centers of the estimated posteriors and the parameterized priors, i.e., the means of distributions. 
Here, we take the obtaining of user embedding $\bm{z}^u$ as an illustration, and Ad embedding $\bm{z}^i$ can be obtained in the same way.
The means of the parameterized priors are used to make up for the unreliable posteriors of the awfully infrequent or new IDs.
\begin{equation}
\begin{aligned}
&\bm{z}^u=g(u)\mu_q(u) + (1-g(u))\mu_p(c(u))\\
\end{aligned}
\end{equation}
As a variant of sigmoid function with the statistics of $u$ as inputs, $g(u)$ is designed to control the weights for $\mu_q^u$ and $\mu_p(u)$ to form the final representation of $u$ during inference:
\begin{equation}
\begin{aligned}
&g(u)=\frac{1}{1+e^{ - \mathcal{F}(u) + \epsilon}}\\
\end{aligned}
\end{equation}
where $\mathcal{F}(u)$ is the accumulated frequency of $u$ in the training dataset and $\epsilon$ is a small constant for numerical stability.
In this way, a certain amount of flexibility is built. The convincing estimated posterior leads the role for a frequent ID, while the convincing prior balances the unreliable posterior for a new or awfully infrequent ID. As shown in Figure \ref{variational-framework}, $\hat{y}$ is then calculated by Equation \ref{predict}.

It is easy to see the advantage of our VELF model. We can effectively make predictions for infrequent and new users and Ads by evaluating posterior and prior distributions end to end. For an infrequent user or Ad, the global knowledge among the similar and frequent users or Ads can be shared during training and inference. The center of the global knowledge is used to represent a new user or Ad to improve the accuracy when inference.

\section{Experiments}
In this section, we conduct experiments with the aim of answering the following three research questions:
\begin{itemize}
    \item[\textbf{RQ1}] How does VELF perform compared to the existing cold-start methods from the perspective of embedding?
    % vs. 非bayesian的退化版本，given normal gaussian priors
    \item[\textbf{RQ2}] How does VELF perform when plugged into various network backbones?
    \item[\textbf{RQ3}] What are the effects of distribution estimate, parameterized and regularized priors in VELF?
\end{itemize}

\subsection{Dataset}
We evaluate the performance of our proposed approaches on three publicly available datasets:
\begin{itemize}[leftmargin=10pt]
\setlength{\itemsep}{-2pt}
\item \textbf{MovieLens-1M\footnote{\url{http://www.grouplens.org/datasets/movielens/}}:} One of the most well-known benchmark dataset. The dataset is made up of 1 million movie ranking instances over thousands of movies and users. The movie ratings are transferred into binary (The ratings at least 4 are turned into 1 and the others are turned into 0).
\item \textbf{Taobao Display Ad Click\footnote{\url{https://tianchi.aliyun.com/dataset/dataDetail?dataId=56}}:} The dataset consists of 26 million ad display / click records generated by 1.14 million users on Taobao website within 8 days.
\item \textbf{CIKM2019 EComm AI\footnote{\url{https://tianchi.aliyun.com/competition/entrance/231721/introduction?lang=en-us}}:} An E-commerce recommendation dataset contains 62 million instances, each of which consists of an item, a user, and a behavior label ('pv,' 'buy,' 'cart,' 'fav'). To match the issue, we convert the instance label to binary (1/0 indicates whether a user has purchased an item or not).
\end{itemize}

We now describe the training sets and test sets preparation.

\textbf{Training sets:} We follow the training setting commonly used in previous research works. For \textbf{MovieLens-1M}, we use the first 80\% instances of each user ordered by time as training set \cite{lee2019melu}
% as in
and further move the users with less than 30 reviews to test set. For \textbf{Taobao Display Ad Click}, the click data that generated in the first 7 days are used as training set, and the data in the last 1 day is the test set in our experiment \cite{zhou2018deep}. For \textbf{CIKM2019 EComm AI}, we use the default training set of this dataset \cite{meta-4}.
    
\textbf{Test sets:} For testing, we prepared 5 different test sets to assess recommendation performance for both new users/items and infrequent users/items. The corresponding definitions for each test set are given in Table~\ref{tab:testset}. Note that the infrequent users/items proportion of the overall users/items is approximately 20\%, which is similar to the definition of the long-tail \cite{meta-2}.

The statistics of the three datasets can be found in Table~\ref{tab:data}, and the utilized features are listed in Table~\ref{tab:testset}. In Table~\ref{tab:testset} and Table~\ref{tab:data}, 'Infreq' is short for 'Infrequent'  and 'fea' is short for 'features'.
The datasets we adopt contain one dataset of online advertising and two datasets of personalized recommendation. Thus we generally call 'Ad' as well as 'Item' as 'Item'.

\begin{table}[]
    \centering\small
    \begin{tabular}{|p{0.008\textwidth}|c|p{0.35\textwidth}|}
    \hline
    % \multirow{6}{a}
    % \multicolumn{3}{|c|}{\textbf{MovieLens-1M}}\\\hline
      \multirow{8}{*}{\rotatebox[origin=c]{90}{\textbf{MovieLens-1M}}}  &Item fea & \textit{title, year of release, genres}\\
        &User fea &  \textit{unique ID, age, gender, occupation}\\\cline{2-3}
        &New user & Users who commented less than 30 times. \\
        &New item & Movies released after 1997.\\
        &Infreq user & 80\% users ordered by the number of posed comment \\
        &Infreq item & 80\% movies ordered by the number of users interacted with it.\\\hline
    \end{tabular}
    \begin{tabular}{|p{0.008\textwidth}|c|p{0.35\textwidth}|}
    \hline
    % \multicolumn{2}{|c|}{\textbf{Taobao Display Ad Click}}\\\hline
        \multirow{11}{*}{\rotatebox[origin=c]{90}{\textbf{Taobao Display Ad Click}}}&Item fea & \textit{Ad ID, category ID, campaign ID, brand ID, Advertiser ID, price}\\
        &User fea & \textit{user ID, Micro group ID, {cms\_group\_id}, gender, age, consumption grade, shopping depth, occupation, city level}\\\cline{2-3}
        &New user & Users who exist only in the default test set. \\
        &New item & Items that exist only in the default test set.\\
        &Infreq user & 60\% users ordered by the number of the number of the related click \\
        &Infreq item & 80\% items ordered by the number of users interacted with it.\\\hline
    \end{tabular}
    \begin{tabular}{|p{0.008\textwidth}|c|p{0.35\textwidth}|}
    \hline
    % \multicolumn{2}{|c|}{\textbf{CIKM2019 EComm AI}}\\\hline
        \multirow{8}{*}{\rotatebox[origin=c]{90}{\textbf{CIKM2019}}}&Item fea & \textit{item ID, category ID, shop ID, brand}\\
        &User fea & \textit{user ID, gender, age, purchasing power}\\\cline{2-3}
        &New user & Users who exist only in the default test set. \\
        &New item & Items that exist only in the default test set.\\
        &Infreq user & 20\% users ordered by the number of the number of the related click.\\
        &Infreq item & 80\% items ordered by the number of users interacted with it.\\\hline
    \end{tabular}
    \caption{Features and Test set construction.}
    \label{tab:testset}
    \vspace{-2.5em}
\end{table}

\begin{table}[]
    \centering\small
    \begin{tabular}{c|r|r|r}
    \toprule
        Dataset & \textbf{MovieLens} & \textbf{Taobao Ad} & \textbf{CIKM2019} \\
        \specialrule{0em}{1pt}{1pt}
        \hline
        \specialrule{0em}{1pt}{1pt}
        \#user & 6,040 & 1,141,729 & 1,050,000 \\
        \#item & 3,706 & 864,811 & 3,934,201 \\
        %\#instance & 1,000,209 & 25,029,435 & 62,428,540 \\
        \#training sample & 630,602 & 21,929,927 & 58,751,493 \\
        \#test (All) & 369,607 & 3,099,508 & 3,677,047 \\
        \#test (New user) & 18,169 & 275,723 & 3,677,047 \\
        \#test (New item) & 196,059 & 87,894 & 114,906 \\
        \#test (Infreq user) & 177,380 & 391,007 & 81,964 \\
        \#test (Infreq item) & 137,508 & 369,561 & 570,590 \\
    \bottomrule
    \end{tabular}
    \caption{Statistics of datasets. \#test stands for the number of instance in different test sets.}
    \label{tab:data}
    \vspace{-2.5em}
\end{table}

\begin{table*}[tp]
  \centering\small
  \caption{Model comparison on three datasets. We record the mean results over 5 runs. Std $\approx$ 0.1\%, extremely statistically significant under unpaired t-test. * indicates the improvement is statistically significant at the significance level of 0.05 over the best baseline on AUC. `Infreq' is short for `Infrequent'.}
  \vspace{-1.0em}
  \label{tab:rq1}
  \begin{tabular}{cc||cc||cc||cc||cc||cc}
    \toprule
    \multirow{10}{*}{\rotatebox{90}{MovieLens-1M}} & \multirow{2}{*}{Methods} & \multicolumn{2}{c||}{New user} & \multicolumn{2}{c||}{New item} & \multicolumn{2}{c||}{Infreq user} & \multicolumn{2}{c||}{Infreq item} & \multicolumn{2}{c}{All} \\
    & & AUC & RelaImpr & AUC & RelaImpr & AUC & RelaImpr & AUC & RelaImpr & AUC & RelaImpr \\
    \specialrule{0em}{1pt}{1pt}
    \cline{2-12}
    \specialrule{0em}{1pt}{1pt}
    & Wide \& Deep & 0.6771 & 5.0\% & 0.6488 & 16.8\% & 0.6955 & 4.7\% & 0.6786 & 4.9\% & 0.7276 & 3.0\%
 \\
    & PNN & 0.6701 & 0.9\% & 0.6470	& 15.4\% & 0.6955 & 4.7\% & 0.6840 & 8.1\% & 0.7275	& 2.9\%
 \\
    & DCN & 0.6785 & 5.9\% & 0.6460	& 14.6\% & 0.6946 & 4.2\% & 0.6778 & 4.5\% & 0.7280	& 3.2\%
 \\
    & xDeepFM & 0.6781 & 5.6\% & 0.6476	& 15.9\% & 0.6958 & 4.8\% & 0.6799 & 5.7\% & 0.7294	& 3.8\%
 \\
    & DeepFM & 0.6686 & 0.0\% & 0.6274 & 0.0\%	& 0.6868 & 0.0\% & 0.6702 & 0.0\% & 0.7210 & 0.0\%
 \\
    & DropoutNet(DeepFM) & 0.6640 & -2.7\% & 0.6298	& 1.9\%	& 0.6875 & 0.4\% & 0.6711 & 0.6\% & 0.7216 & 0.3\%
 \\
    & MWUF(DeepFM) & 0.6712	& 1.5\%	 & 0.6573 & 23.5\% & 0.6991	& 6.6\%	& 0.6886 & 10.8\% & 0.7342 & 6.0\%
 \\
    & VELF(DeepFM) & $\textbf{0.7112}^*$	& $\textbf{25.3\%}$ & $\textbf{0.7106}^*$ & $\textbf{65.3\%}$ & $\textbf{0.7117}^*$	& $\textbf{13.3\%}$ & $\textbf{0.7009}^*$ & $\textbf{18.0\%}$ & $\textbf{0.7551}^*$ & $\textbf{15.4\%}$
 \\
    \midrule
    \multirow{10}{*}{\rotatebox{90}{Taobao Display Ad Click}} & \multirow{2}{*}{Methods} & \multicolumn{2}{c||}{New user} & \multicolumn{2}{c||}{New item} & \multicolumn{2}{c||}{Infreq user} & \multicolumn{2}{c||}{Infreq item} & \multicolumn{2}{c}{All} \\
    & & AUC & RelaImpr & AUC & RelaImpr & AUC & RelaImpr & AUC & RelaImpr & AUC & RelaImpr \\
    \specialrule{0em}{1pt}{1pt}
    \cline{2-12}
    \specialrule{0em}{1pt}{1pt}
    & Wide \& Deep & 0.5573	& -30.1\% & 0.5995 & -8.3\%	& 0.5713 & -18.2\% & 0.5964	 & -12.5\% & 0.6204 & -8.1\%
 \\
    & PNN & 0.5535 & -34.8\% & 0.6064 & -1.9\% & 0.5547	& -37.3\% & 0.5991 & -10.1\% & 0.6140 & -13.0\%
 \\
    & DCN & 0.5843 & 2.8\% & 0.6141	& 5.2\%	& 0.5873 & 0.1\% & 0.6157 & 5.0\% & 0.6256 & -4.1\%
 \\
    & xDeepFM & 0.5831 & 1.3\% & 0.6104	& 1.8\%	& 0.5874 & 0.2\% & 0.6129 & 2.5\% & 0.6328 & 1.4\%
 \\
    & DeepFM & 0.5820 & 0.0\% & 0.6085 & 0.0\% & 0.5872	& 0.0\% & 0.6102 & 0.0\% & 0.6310 & 0.0\%
 \\
    & DropoutNet(DeepFM) & 0.5848 & 3.4\% & 0.6179 & 8.7\% & 0.5884 & 1.4\% & 0.6303 & 18.2\% & 0.6340 & 2.3\%
 \\
    & MWUF(DeepFM) & 0.5819 & -0.1\% & 0.6117 & 2.9\% & 0.5896 & 2.8\% & 0.6244 & 12.9\% & 0.6322 & 0.9\%
 \\
    & VELF(DeepFM) & $\textbf{0.5895}^*$ & $\textbf{9.1\%}$ & $\textbf{0.6220}^*$ & $\textbf{12.4\%}$ & $\textbf{0.5998}^*$ & $\textbf{14.4\%}$ & $\textbf{0.6332}^*$ & $\textbf{20.9\%}$ & $\textbf{0.6394}^*$ & $\textbf{6.4\%}$
 \\
    \midrule
    \multirow{10}{*}{\rotatebox{90}{CIKM2019 Ecomm AI}} & \multirow{2}{*}{Methods} & \multicolumn{2}{c||}{New user} & \multicolumn{2}{c||}{New item} & \multicolumn{2}{c||}{Infreq user} & \multicolumn{2}{c||}{Infreq item} & \multicolumn{2}{c}{All} \\
    & & AUC & RelaImpr & AUC & RelaImpr & AUC & RelaImpr & AUC & RelaImpr & AUC & RelaImpr \\
    \specialrule{0em}{1pt}{1pt}
    \cline{2-12}
    \specialrule{0em}{1pt}{1pt}
    & Wide \& Deep & 0.7467 & 0.0\% & 0.6877 & 0.1\% & 0.7451 & 0.3\% & 0.7139 & 0.3\% & 0.7467 & 0.0\%
 \\
    & PNN & 0.7468 & 0.0\% & 0.6882 & 0.3\% & 0.7433 & -0.4\% & 0.7145 & 0.6\% & 0.7468 & 0.0\%
 \\
    & DCN & 0.7468 & 0.0\% & 0.6883 & 0.4\% & 0.7449 & 0.2\% & 0.7143 & 0.5\% & 0.7468 & 0.0\%
 \\
    & xDeepFM & 0.7464 & -0.1\% & 0.6867 & -0.5\% & 0.7438 & -0.2\% & 0.7132 & 0.0\% & 0.7464 & -0.1\%
 \\
    & DeepFM & 0.7467 & 0.0\% & 0.6876 & 0.0\% & 0.7443 & 0.0\% & 0.7132 & 0.0\% & 0.7467 & 0.0\%
 \\
    & DropoutNet(DeepFM) & 0.7467 & 0.0\% & 0.6886 & 0.5\% & 0.7455 & 0.5\% & 0.7138 & 0.3\% & 0.7467 & 0.0\%
 \\
    & MWUF(DeepFM) & 0.7483 & 0.6\% & 0.6887 & 0.6\% & 0.7450 & 0.3\% & 0.7164 & 1.5\% & 0.7483 & 0.6\%
 \\
    & VELF(DeepFM) & $\textbf{0.7497}$ & $\textbf{1.2\%}$ & $\textbf{0.6967}^*$ & $\textbf{4.9\%}$ & $\textbf{0.7492}$ & $\textbf{2.0\%}$ & $\textbf{0.7228}^*$ & $\textbf{4.5\%}$ & $\textbf{0.7497}$ & $\textbf{1.2\%}$
 \\
    \bottomrule
  \end{tabular}
\vspace{-1em}
  
\end{table*}

\begin{figure*}[t]
\captionsetup{justification=centering}
\centering
    \begin{minipage}[t]{0.44\linewidth}
      \centering
      \centerline{\includegraphics[width=1\linewidth]{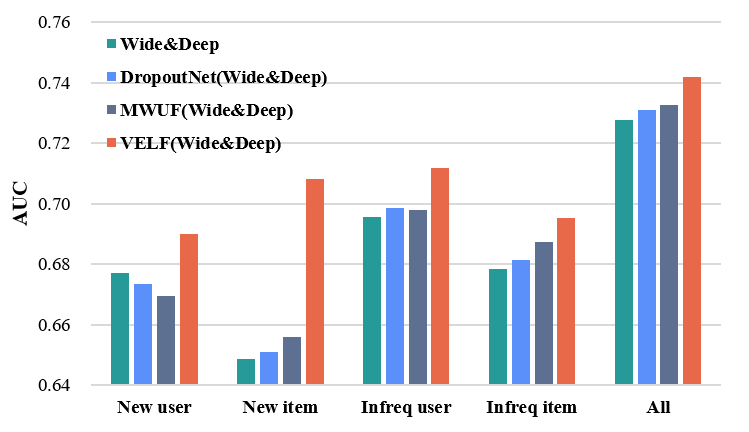}}
      \vspace{-0.5em}
      \centerline{\footnotesize{(a)}}
       \vspace{-0.5em}
      \centering
    \end{minipage}%
    \hspace{0.2em}
    \begin{minipage}[t]{0.44\linewidth}
      \centering
      \centerline{\includegraphics[width=1\linewidth]{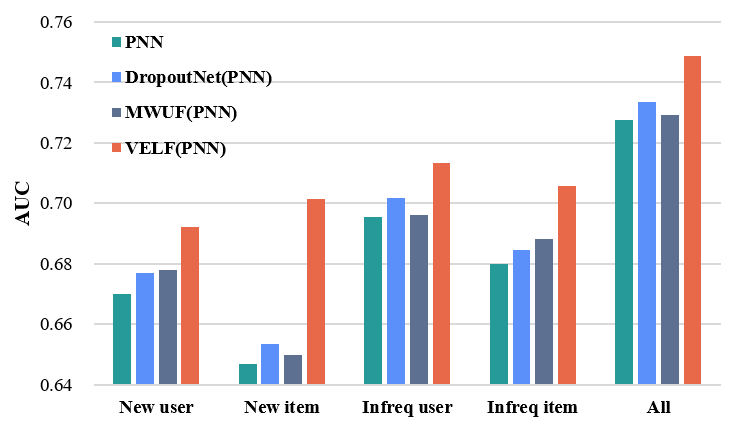}}
      \vspace{-0.5em}
      \centerline{\footnotesize{(b)}}
       \vspace{-0.5em}
      \centering
    \end{minipage}%
    
    \begin{minipage}[t]{0.44\linewidth}
      \centering
      \centerline{\includegraphics[width=1\linewidth]{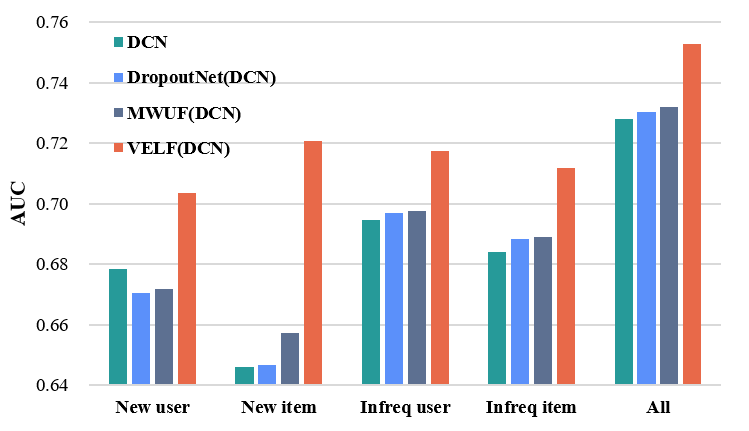}}
      \vspace{-0.5em}
      \centerline{\footnotesize{(c)}}
       \vspace{-0.5em}
      \centering
    \end{minipage}%
    \hspace{0.2em}
    \begin{minipage}[t]{0.44\linewidth}
      \centering
      \centerline{\includegraphics[width=1\linewidth]{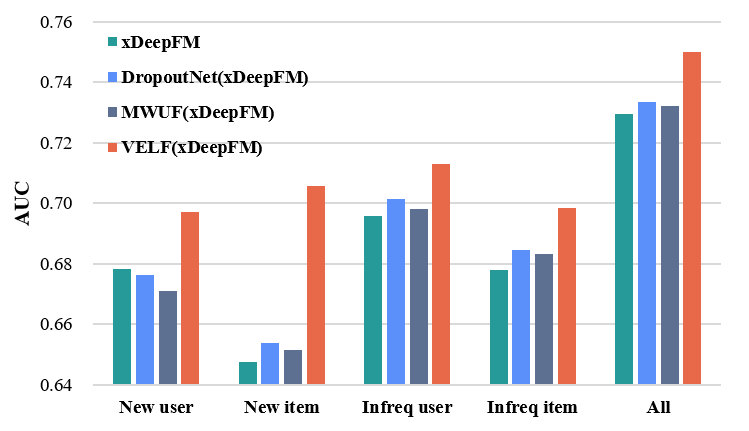}}
      \vspace{-0.5em}
      \centerline{\footnotesize{(d)}}
       \vspace{-0.5em}
      \centering
    \end{minipage}%
\vspace{-0.5em}
\caption{Performence on MovieLens-1M over four popular feature-crossing backbones: \\(a) Wide\&Deep, (b) PNN, (c) DCN, (d) xDeepFM. }
\label{rq2}
%\vspace{-1.0em}
\end{figure*}

\subsection{Baselines}
We divide our baselines into two groups based on their methods. 

The first group contains state-of-the-art approaches for dealing with the cold-start problem.
(1) DropoutNet~\cite{volkovs2017dropoutnet} is a famous cold-start approach that uses the average representations of interacting items/users to improve user/item representations.
(2) MWUF~\cite{zhu2021learning} introduce Meta Scaling and Shifting Networks to construct scaling and shifting functions for each item, with the scaling function directly transforming cold item ID embeddings into warm feature space and the shifting function producing stable embeddings from noisy embeddings.

The common Feature-Crossing techniques developed for overall recommendation are included in the second group. The second group serves as baselines without cold-start alleviating component as well as backbones to test the generalization and adaptability of our proposed VELF.
(1) DeepFM~\cite{guo2017deepfm} is a deep recommendation method that learns both low- and high-level interactions between fields. 
(2) Wide\&Deep~\cite{cheng2016wide} develop wide linear models and deep neural networks together to enhance their respective abilities.
(3) DCN~\cite{wang2017deep}, which is based on DNN, explicitly applies feature crossing at each layer, eliminating the need for human feature engineering.
(4) xDeepFM~\cite{lian2018xdeepfm} generates feature interactions directly at the vector-wise level, allowing it to learn specific bounded-degree feature interactions explicitly at the low- and high-order levels.
(5) PNN~\cite{qu2016product} employs a feature extractor to investigate feature interactions among inter-field categories.

\subsection{Experimental Settings}
% \subsubsection{Data Preparation}

\subsubsection{Implementation Details}
We utilize the same model settings for all approaches on each dataset to provide a fair comparison. For all the three datasets, we fix embedding size as 8 and DNN as 3 FC layers with 200 hidden units. Furthermore, for xDeepFM and DCN, we set the number of cross layer to 2. We optimize all approaches using mini-batch Adam, where the learning rate is searched from \{1e-5, 5e-4, 1e-4, ..., 1e-2\}. Furthermore, the batch size of all models is set to 256 for the MovieLens-1M dataset and 4096 for others. 
% \textcolor{red}{setting for feature crossing baselines,eg., crossing layer num}

\subsubsection{Evaluation Metrics}
AUC \cite{auc} is a common metric for both recommendation \cite{zhu2021learning} and advertising \cite{zhou2018deep}. It measures the goodness of order by ranking all the items with prediction. Thus following the cold-start work \cite{zhu2021learning,meta-2}, we adopt AUC as the main metric. In addition, we follow \cite{zhu2021learning,zhou2018deep} to introduce RelaImpr metric to measure relative improvement over models. For a random guesser, the value of AUC is 0.5. Hence, RelaImpr is defined as:
\begin{equation}
\setlength{\abovedisplayskip}{1pt}
\rm RelaImpr=(\frac{AUC(measured\ model)-0.5}{AUC(base\ model)-0.5}-1) \times 100\%
\setlength{\belowdisplayskip}{1pt}
\end{equation}

\subsection{Comparison with State-of-the-arts (RQ1)}
We compare our VELF with the SOTA methods to alleviate cold-start problem from the perspective of embedding learning, i.e., DropoutNet \cite{volkovs2017dropoutnet} and MWUF \cite{zhu2021learning}. 
Comparisons with state-of-the-arts are conducted on DeepFM \cite{guo2017deepfm} which is one of the most popular model structures used in industry.
For more detailed and directed analysis, we also report the results of the second group baseline models mentioned before. Evaluations are conducted on three benchmark datasets to report the mean results over five runs. The results are shown in Table \ref{tab:rq1}.

\textbf{The effectiveness of VELF.} 
VELF outperforms all the baselines on three datasets. Especially, the AUC improvements on 'New' and 'Infreq' test datasets are much more remarkable than 'All' test datasets. The results confirm the effectiveness of VELF to alleviate cold-start problem in CTR prediction. 

\textbf{Discussions.} 
Firstly, with VELF, and similarly with DropoutNet and MWUF, the AUC improvements on 'Item' datasets are more significant than those on 'User' datasets. 
%This indicates that VELF works better on such scenarios that whose IDs have natural cluster, eg., the similarity among items in the same category is much stronger and more interpretable than that among users with the same gender. 
The reason is that in these three datasets, the data sparsity problem of items is more severe than that of users.
Secondly, with DropoutNet, the performences on MovieLens datasets are much weaker than those on Taobao Display Ad and CIKM2019. To explain, recall that Taobao Display Ad and CIKM2019 have more abundant attributes for users and items which are relatively limited in MovieLens. Thus, content-based methods are more sensitive to the limitation of side information.
Thirdly, on CIKM2019 dataset, AUC results of 'New user' and 'All' are equal. The reason is that users in test dataset and training dataset are non-overlapping by the default splitting settings.

\subsection{Generalization Experiments (RQ2)}
In Section 4.4, we evaluate the effectiveness of VELF on the backbone of DeepFM. To further evaluate the generalization of VELF, we compare VELF with DropoutNet and MWUF in another four different popular network backbones, including Wide\&Deep, PNN, DCN and xDeepFM. Experiments are conducted on the MovieLens-1M dataset.

The same as DropoutNet and MWUF, our VELF can be adapted to any network backbones by replacing the user and item embedding module. The experimental results are reported in Figure \ref{rq2}. The results indicate that VELF can constantly achieve the best performance with various base models.

\subsection{Ablation Study (RQ3)}
In this section, we demonstrate the advantages of our proposed distribution estimate, parameterized and regularized priors in VELF. We present an ablation study on MovieLens-1M dataset by evaluating several models based on DeepFM which is one of the lightest structures: (1) VELF: the overall framework; (2) VELF(Point): degenerate distribution estimate into point estimate by directly adopting $\mu_q^u$ as $\bm{z}^u$ and $\mu_q^i$ as $\bm{z}^i$; (3) VELF(No-R): degenerate the parameterized and regularized priors into one layer parameterized priors without regularization by only keeping $p_{\bm{\phi}_p^u}(\bm{z}^u)$ and $p_{\bm{\phi}_p^i}(\bm{z}^i)$; (4) VELF(Fixed): degenerate the parameterized and regularized priors into one layer fixed normal priors. 
The mean AUC results over 5 runs are reported in Table \ref{tab:rq3}. 

Firstly, according to Table \ref{tab:rq1} and Table \ref{tab:rq3}, VELF(Point) does not outperform DropoutNet much. This indicates the superiority of distribution estimate over point estimate. Secondly, VELF(No-R) outperforms VELF(Fixed) which confirms our claim that our proposed parameterized priors can improve the generalization ability. Thirdly, VELF is more effective than VELF(No-R). This demonstrates that constraining the parameterized priors to be close to a normal hyper-prior is helpful to further improve the generalization ability.

\begin{table}[tp]
  \centering\small
  \caption{Ablation study. The averaged AUC results over 5 runs are reported. Std $\approx$ 0.1\%, extremely statistically significant under unpaired t-test.}
  \vspace{-1.0em}
  \label{tab:rq3}
  \begin{tabular}{c|c|c|c|c|c}
    \toprule
    Methods & New user & New item & Infreq user & Infreq item & All \\
    \midrule
    VELF & 0.7112 & 0.7106 & 0.7117 & 0.7009 & 0.7551 \\
    VELF(No-R) & 0.6843 & 0.7037 & 0.7058 & 0.6935 & 0.7502 \\
    VELF(Fixed) & 0.6723 & 0.6831 & 0.7070 & 0.6871 & 0.7402 \\
    VELF(Point) & 0.6568 & 0.6508 & 0.6869 & 0.6723 & 0.731 \\
    \bottomrule
  \end{tabular}
  \vspace{-1.5em}
\end{table}

\section{Conclusion}
In this paper, we propose a general Variational Embedding Learning Framework (VELF) to improve the generalization ability and robustness of embedding learning for cold-start users and Ads. VELF regards the embedding learning as a distribution estimate process, which means that embeddings are inferred from a series of shared distributions based on Bayesian inference. Thus the embeddings, especially the cold-start ones, can benefit from statistical strength among all the users and Ads. 
Besides, we develop a parameterized and regularized prior mechanism which can naturally utilizing the rich side information to further suppress overfitting. 
The embedding distributions and the discriminative CTR predicition network parameters are learned end-to-end without strict requirements on extra training data or training stages.
Experiments on several recommendation tasks show that CTR models with VELF can achieve better performances. Our future work will include the interactively modeling of user and Ad based on VELF and specific feature-crossing techniques under VELF.

\bibliographystyle{ACM-Reference-Format}
\bibliography{main}

\appendix

\section{Infra-cost Analysis}
We conducted experiments on a machine with a 10-core Intel(R) Xeon(R) CPU E5-2698 v4 @ 2.20GHz and 100G memory, using tensorflow2.2.0. On Movielens dataset, 1.3G of memory was required by DeepFM when 1.4G of memory for DeepFM+VELF. The total computational complexity of DeepFM+VELF is less than 2x. DeepFM. Thanks to the high parallelism of CPU and tensorflow, the whole training time of DeepFM+VELF was only extended 10\% compared to DeepFM under the same hardware constraints.

\end{document}